\input{epsf}
\documentstyle[prl,aps]{revtex}
\begin{document}
%\draft

\title{Evolving test-fields in a black-hole geometry}

\author{Nils Andersson}

\address{Department of Physics, Washington University, 
St Louis MO 63130, USA}

\twocolumn[

\maketitle

\begin{abstract}
\widetext
We consider the initial value problem for a massless scalar field in
the Schwarzschild geometry. When constructed using a complex-frequency
approach the necessary Green's function splits into three components.
We discuss all of these in some detail: 1) The contribution from the
singularities (the quasinormal modes of the black hole) is approximated
and the mode-sum is demonstrated to converge after a certain well
defined time in the evolution. A dynamic description of the
mode-excitation is introduced and tested. 2) It is shown how a
straightforward low-frequency approximation to the integral along the
branch cut in the black-hole Green's function leads to the anticipated
power-law fall off at very late times. We also calculate higher order
corrections to this tail and show that they provide an important
complement to the leading order.  3) The high-frequency problem is also
considered. We demonstrate that the combination of the obtained
approximations for the quasinormal modes and the power-law tail provide
a complete description of the evolution at late times.  Problems that
arise (in the complex-frequency picture) for early times are also
discussed, as is the fact that many of the presented results generalize
to, for example, Kerr black holes.
\end{abstract}
]


\narrowtext

\section{Cauchy's problem for perturbed black holes}

This paper concerns the evolution of a test-field (be it scalar,
electromagnetic or a perturbation of the gravitational field itself) in
a spacetime that contains a black hole. That is, we consider the
problem that is associated with Cauchy in the framework of general
relativity. Because of the inherent nonlinearity of Einstein's theory
this problem is generally not amenable to analytic calculations. But if
the wave-field is sufficiently weak that its contribution to the
spacetime curvature can be neglected the evolution equations reduce to
a wave equation with a complicated effective potential.  This is the
realm of black-hole perturbation theory \cite{cbook,nbook} in which the
initial-value problem can be approached by ``standard'' methods
\cite{leaverprd,andersson95}.  The purpose of the present work is to
contribute a more detailed understanding of the many intricacies
associated with the evolution of a weak wave-field in a black-hole
geometry.

One can argue that this kind of discussion is of little importance to
physics. It may seem obvious that much relevant information will be
lost when the equations of general relativity are linearised. But it
turns out that the perturbation approach provides surprisingly accurate
results in many situations. An interesting example of this is the case
of two colliding black holes \cite{price_pullin}.  This obviously does
not mean that the linear equations render a fully nonlinear approach
useless. It would be truly surprising if no new phenomena were to be
unveiled by detailed nonlinear calculations, but linear studies provide
important benchmarks against which such fully nonlinear, numerical
calculations can (and should) be tested. Also --- and of equal
importance --- is the fact that the linear problem can be approached
``analytically''. This can lead to an improved understanding of the
underlying physics and information that can be extremely difficult to
infer from purely numerical data.

The problem we consider here is in many ways an old one. The evolution
of a test-field in a black-hole background was first considered more
than 25 years ago \cite{vishu}. The general features of such an
evolution are well-known \cite{nbook}.  The ``response'' of the black
hole --- as seen by a distant observer --- can be divided into three
components.  Radiation emitted ``directly'' by the perturbation source
will dominate at early times. This radiation  depends on the exact
character of the initial field. In contrast, the late-time response
depends mainly on the parameters of the black-hole. The exponentially
damped oscillations of the black-hole quasinormal modes carry a
considerable part of the total radiated energy in many astrophysical
processes (such as gravitational collapse)
\cite{cpm78,cpm79,seidel90,seidel91}.  Finally, the wave field falls
off with time according to a power law at very late times
\cite{price1,price2}.

The initial-value problem for black-hole perturbations have been
considered by several authors. In an impressive study, Leaver
\cite{leaverprd} discussed both the excitation of quasinormal modes and
the nature of the power-law tails. The quasinormal-mode problem was
later considered by Sun and Price \cite{sun_price} and also by the
present author \cite{andersson95}.  %Several relevant studies of
quasinormal modes in %other radiating systems have emerged from the
physics %group in Hong Kong [REF].  Late-time tails have recently been
studied by Gundlach, Price and Pullin \cite{gundlach1,gundlach2} and
Ching {\em et al.} \cite{ching}.
 
Even though the problem is far from new, there are several reasons why
it need be investigated further. Although the response of a black hole
to an impinging wavepacket will almost exclusively be dominated by the
slowest damped quasinormal modes --- and present methods can reliably
account for the excitation of these modes \cite{leaverprd,andersson95}
--- several questions remain. For example: What is the role of the
highly damped modes? It is known that an infinite number of quasinormal
modes exist for each radiative multipole $\ell$
\cite{nollert,andersson93}, but our understanding of the role of the
higher overtones is rather poor.  In fact, it is not at all clear
whether the mode-sum is convergent or not \cite{leaverprd}.  Our
understanding of the power-law tail is also somewhat unsatisfactory.
The leading behaviour has been calculated in different ways
\cite{leaverprd,ching}, but the resultant formulae are only truly
useful at very late times. In a typical evolution scenario there is a
considerable time-window in which the signal is no longer dominated by
the quasinormal modes, but the leading order power-law tail has not yet
taken over.  Is it possible to derive a ``higher-order'' tail
expression that describes the evolution adequately for the intermediate
times?  These questions (and several others) are addressed in the
present paper.

\section{The problem and a formal solution}

\subsection{A massless scalar field in the Schwarzschild geometry}

In order to make the presentation clear we have chosen to specialize
this investigation to the case of a massless scalar field and
Schwarzschild black holes. This is, of course, a model problem since no
scalar fields have yet been observed in Nature. But this does not mean
that our results are of restricted value.  On the contrary:  Because
the equations that govern other perturbing fields (such as an
electromagnetic testfield or a gravitational perturbation of the
metric) are similar to the one for a scalar field \cite{nbook}, the
results presented here are easily extended to all other relevant
cases.  Furthermore, as will be discussed in section VI, it seems
likely that many of the present results can  be adapted also to the
case of rotating black holes.

In the background geometry of a Schwarzschild black hole 
a massless scalar field evolves according to
\begin{equation}
\Box \Phi = 0 \ .
\end{equation}
Because of the underlying spherical symmetry it is meaningful to 
introduce the decomposition
\begin{equation}
\Phi_{\ell m} =  {u_\ell (r_\ast,t) \over r} Y_{\ell m} (\theta, \varphi) 
\ ,
\end{equation}
where $Y_{\ell m}$ are the standard spherical harmonics. 
The function 
$u_\ell(r_\ast,t)$ then solves the wave equation
\begin{equation}
\left[ {\partial^2 \over \partial r_\ast^2}- 
{\partial^2 \over \partial t^2} - V_\ell(r) \right] 
u_\ell  = 0 \ ,
\label{waveq}\end{equation}
where the effective potential is
\begin{equation}
V_\ell (r)  = \left(1 -{2M\over r} \right) \left[ {\ell(\ell+1) 
\over r^2} + {2M\over r^3} \right] \ ,
\label{vpot}\end{equation}
and $M$ is the mass of the black hole (we use geometrized units
$c=G=1$). 
The ``tortoise'' coordinate $r_\ast$ is defined by
\begin{equation}
{d \over dr_\ast} = \left( 1 - {2M\over r} \right) {d \over dr}  \ .
\end{equation}

Let us now suppose that we are given a specific scalar field at some
time (we will use $t=0$), and that we want to deduce the future
evolution of this field.  That is, we require a scheme for calculating
(for each $\ell$) $u_\ell(r_\ast,t)$ once we are given
$u_\ell(r_\ast,0)$ and $\partial_t u_\ell(r_\ast,0)$. This problem is
typically approached via a Green's function.

\subsection{The black-hole Green's function}

It is well-known  that the 
time-evolution of a wave-field $u_\ell(r_\ast,t)$
follows immediately from
\begin{eqnarray}
u_\ell(r_\ast,t) &=& \int G(r_\ast,y,t)\partial_t 
u_\ell(y,0) dy \nonumber \\
&+&  \int \partial_t G(r_\ast,y,t) u_\ell(y,0) dy \ ,
\label{Gevol}\end{eqnarray}
for $t>0$ (we will discuss the appropriate limits of integration in section IIIC). 
The (retarded) Green's function is defined by
\begin{equation}
\left[ {\partial^2 \over \partial r_\ast^2}- 
{\partial^2 \over \partial t^2} - V_\ell(r) \right] 
G(r_\ast,y,t) = \delta(t)\delta(r_\ast-y) \ ,
\label{Geqn}\end{equation}
together with the condition $G(r_\ast,y,t) = 0$ for $t\le 0$.

To find the Green's function $G(r_\ast,y,t) $ is our main task.
Once we know it, we can study the evolution of any initial field
by evaluating the integrals in (\ref{Gevol}). 
 
The first step in finding  $G(r_\ast,y,t)$ consists of reducing 
(\ref{Geqn}) to
an ordinary differential equation. To do this we use the 
integral transform \cite{andersson95}
\begin{equation}
\hat{G}(r_\ast,y,\omega) = \int_{0^-}^{+\infty}
G(r_\ast,y,t) e^{i\omega t} dt \ .
\label{ftrafo}\end{equation}
This transform is well defined as long as ${\rm Im}\ \omega \ge 0$,
and the corresponding inversion formula is
\begin{equation} 
G(r_\ast,y,t) = {1 \over 2\pi} \int_{-\infty+ic}^{+\infty+ic} 
\hat{G}(r_\ast,y,\omega) e^{-i\omega t} d\omega \ ,
\label{greent}\end{equation}
where $c$ is some positive number.

The Green's function $\hat{G}(r_\ast,y,\omega)$ can now be expressed in 
terms of
two linearly independent solutions to the homogeneous equation
\begin{equation}
\left[ {d^2 \over d r_\ast^2} + \omega^2 - V_\ell(r) \right] 
\hat{u}_\ell(r_\ast,\omega) = 0 \ .
\label{rweq}\end{equation}
The two required solutions are defined by their asymptotic behaviour.
The first solution corresponds to purely ingoing waves crossing the
event horizon:
\begin{equation}
\hat{u}_\ell^{\rm in}(r_\ast,\omega) \sim \left\{ \begin{array}{ll}
e^{ -i\omega  r_\ast}\ , & r_\ast\rightarrow  -\infty \ , \\
A_{\rm out}(\omega)e^{ i\omega  r_\ast}+A_{\rm in} (\omega) 
e^{ -i\omega  r_\ast}\ , &r_\ast \rightarrow   
+\infty \ ,
\end{array}
\right.
\label{inmode}\end{equation}
and the second solution behaves as a purely outgoing wave at spatial 
infinity:
\begin{equation}
\hat{u}_\ell^{\rm up}(r_\ast,\omega) \sim \left\{ \begin{array}{ll}
B_{\rm out}(\omega)e^{ i\omega  r_\ast}+B_{\rm in}(\omega)e^{ -i\omega  
r_\ast}\ ,
&r_\ast \rightarrow   
-\infty \ ,  \\
e^{ +i\omega  r_\ast}\ , & r_\ast\rightarrow  +\infty \ .
\end{array}
\right.
\label{upmode}\end{equation}

Using these two solutions the Green's function can be written
\begin{equation}
\hat{G}(r_\ast,y,\omega) = -{ 1 \over 2i\omega A_{\rm in} (\omega)} 
\left\{ \begin{array}{lll} 
 \hat{u}_\ell^{\rm in}(r_\ast,\omega) \hat{u}_\ell^{\rm up}(y,\omega) 
 \ , & r_\ast < y \ , \\ \\
\hat{u}_\ell^{\rm in}(y,\omega) \hat{u}_\ell^{\rm up}(r_\ast,\omega) 
 \ , & r_\ast > y \ .
\label{greenf}\end{array} \right.
\end{equation}
Here we have used the Wronskian relation 
\begin{equation}
W(\omega) \equiv \hat{u}_\ell^{\rm in} {d \hat{u}_\ell^{\rm u p} \over dr_\ast}
- \hat{u}_\ell^{\rm up} {d \hat{u}_\ell^{\rm in} \over dr_\ast}
= 2i\omega A_{\rm in} (\omega) \ . 
\label{wronsk}\end{equation}

\subsection{Using complex frequencies}

The problem can now, in principle, be approached by direct numerical
integration of (\ref{rweq}) for (almost) real values of $\omega$ and
subsequent inversion of (\ref{greent}). This approach should lead to
reliable results and an accurate representation of the evolution, as
long as some obvious care is taken in each step. A multitude of
examples of this approach can be found in work relating to particles
orbiting black holes (see \cite{nbook} for an exhaustive list of
references). For the evolution of a test-field, when we want to explain
why different features seen in the emerging waves arise, it may be
useful to follow an alternative route, however.  An approach that often
proves useful when one wants to isolate the behaviour of a Green's
function in different time intervals is based on bending the
integration contour in (\ref{greent}) into the lower half of the
complex $\omega$-plane. This is the approach that we will follow here.

What do we expect to learn by analytically continuing the Green's
function in this way? First of all, it is well known that
$\hat{G}(r_\ast,y,\omega) $ has an infinite number of distinct
singularities in the lower half of the $\omega$-plane. These correspond
to the black-hole quasinormal modes and occur at frequencies for which
the Wronskian $W(w)$ vanishes. That is, for a quasinormal mode the two
solutions  $\hat{u}_\ell^{\rm in}$ and $\hat{u}_\ell^{\rm up}$ are
linearly dependent. To determine the quasinormal-mode frequencies is
not a trivial task, but several accurate methods have been devised
\cite{leaverprs,andersson92,al,ns}. The mode frequencies do not,
however, contain all the information that is required to evaluate the
Green's function.  While it is formally straightforward to use the
residue theorem to determine the mode-contribution it is, in practice,
non-trivial to evaluate the resultant expressions. One must be able to
approximate the eigenfunction associated with each quasinormal mode.

\begin{figure}
\centerline{\epsfxsize=250pt \epsfbox{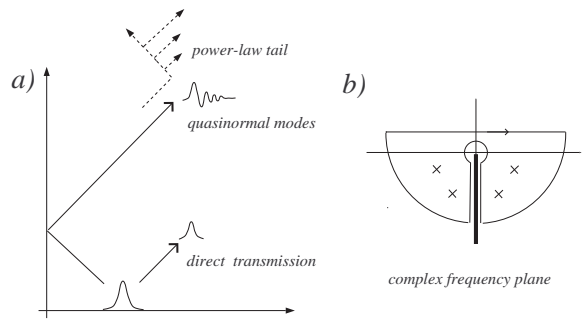}}
\caption{ a) Schematic description of a black holes response to initial
data of compact support. The directly transmitted wave (from a source
point $y$) arrives at a distant observer (at $r_\ast$) roughly at
$t-r_\ast+y=0$. The black holes response, that is dominated by
quasinormal mode ringing, reaches the observer at roughly
$t-r_\ast-y=0$. At very late times the signal falls off as an inverse
power of time. This power-law tail arises because of multiple
backscattering off the spacetime curvature. b) Integration contours in
the complex frequency plane. The original inversion contour for the
 Green's function lies above the real frequency axis. When analytically
continued in the complex plane this contour can be replaced by the sum
of 1) the quasinormal modes [the singularities of
$\hat{G}(r_\ast,y,\omega)$; the first few are represented by crosses in
the figure] 2) an integral along the branch cut (a thick line along the
negative imaginary $\omega$ axis in the figure), that leads to the
power law tail, and 3) high frequency arcs (that one would expect
vanish at most times, but they should also lead to rougly ``flat space
propagators'' at early times). }
\label{fscheme}\end{figure}

In the complex-frequency picture the late-time power-law tail is
associated with the existence of a branch cut in $\hat{u}_\ell^{\rm
up}$.  This cut is usually placed along the negative imaginary
$\omega$-axis.  It has been demonstrated that the behaviour at very
late times can be obtained from a low-frequency approximation of the
integral along the branch cut \cite{leaverprd}. As regards the
radiation that reaches an observer more or less directly from the
source it has been suggested \cite{leaverprd} that it can be associated
with the large-frequency arcs that are required to ``close the
contour'' in the complex $\omega$-plane (see Figure~\ref{fscheme}).
One can argue that this should be the case in a handwaving way: For
large frequencies the Green's function limits to the familiar
flat-space propagator \cite{leaverprd}. As yet there are no detailed
studies of the high-frequency problem, however.

\subsection{The asymptotic approximation}

In this paper we want to pursue the problem analytically as far as
possible.  This means that we will often prefer a simplifying
approximation over a less transparent numerical calculation.  The hope
is that this will lead to a reasonably accurate description of the
evolution, and at the same time provide better insight into the
underlying physics.  Once one has acquired this understanding it will
be meaningful to perform a more accurate analysis.

In this context, a useful approximation follows if one assumes that
spacetime is essentially flat in the region of both the observer and
the initial data (that should be of compact support).  We consequently
assume that i) the observer is situated far away from the black hole.
This means that $r_\ast/M>>1$ in (\ref{greenf}).  ii) the initial data
has considerable support only far away from the black hole. This
implies that only the region where $y/M>>1$ contributes significantly
to (\ref{Gevol}).

To make life easier we will also assume that the initial
data has no support outside the observer (only $y<r_\ast$ are relevant). With all these restrictions
the frequency-domain Green's function (\ref{greenf}) simplifies to
\begin{equation}
\hat{G}(r_\ast,y,\omega) \approx -{ 1 \over 2i\omega} 
\left[ e^{i\omega(r_\ast-y)} + { A_{\rm out} \over A_{\rm in} }
e^{i\omega(r_\ast+y)} \right]  \ . 
\label{greenapp}\end{equation}
In the following we will refer to this as the ``asymptotic approximation''
since it follows when we use the large-argument asymptotics for
$\hat{u}_\ell^{\rm in}$ and $\hat{u}_\ell^{\rm up}$ in (\ref{greenf}). 
The usefulness of this approximation should be obvious. 

\section{Quasinormal modes}

\subsection{Mode-contribution to the Green's function}

As already mentioned, the quasinormal modes correspond to complex
frequencies $(\omega_n)$ for which the Wronskian $W(w)$ vanishes. This
means that  $A_{\rm in}(\omega_n) = 0$ and consequently it is useful to
define a quantity $\alpha_n$ by
\begin{equation}
A_{\rm in}(\omega) \approx (\omega -\omega_n) \alpha_n \ ,
\label{alph}\end{equation}
in the vicinity of the mode. 
Then it follows from the residue-theorem (and the fact
that modes in the third and fourth quadrant are in one-to-one
correspondence, see Figure 4.4 in \cite{nbook}) 
that the total contribution from the 
modes to the time-domain Green's function can be written \cite{andersson95} 
\begin{equation}
G^Q(r_\ast,y,t) =  {\rm Re} \left[ \sum_{n=0}^\infty 
B_n e^{-i\omega_n ( t-r_\ast-y)} 
\right] \ .
\label{Gmodes}\end{equation}
Here we have defined
\begin{equation}
B_n = {A_{\rm out} (\omega_n)\over \omega_n \alpha_n }  \ .
\end{equation}
We have also used the asymptotic approximation (\ref{greenapp}), and
the sum is over all quasinormal modes in the fourth quadrant.  That
this expression provides an accurate representation of the
mode-excitation has already been demonstrated
\cite{leaverprd,sun_price,andersson95}.  Typical results obtained using
(\ref{Gmodes}) are shown in Figures 3a-c in \cite{andersson95}.

\subsection{Convergence of the quasinormal-mode sum}

When one evolves a test-field in the Schwarzschild geometry one
typically finds that the response of the black hole that is associated
with the data at $y$ [cf. (\ref{Gevol})] reaches the observer  roughly
when $t-r_\ast-y=0$. This is not too surprising: The slowest damped
quasinormal modes can be associated with the peak of the effective
potential \cite{sw}. Hence, one would expect the response to follow
once the specific part of the initial data has had time to reach the
peak of the potential (roughly at $r_\ast = 0$) and then travel back to
the observer.

An obvious question concerns the convergence of the quasinormal-mode
sum (\ref{Gmodes}). At what times (if any) will the sum be convergent?
Previous evidence for gravitational perturbations  and the first seven
modes \cite{leaverprd} suggests that the mode sum is convergent at late
times, but fails to provide a lower limit of $t$ at which this
convergence starts. As far a late times are concerned it is easy to
convince oneself that the sum should converge: Two consecutive terms in
(\ref{Gmodes}) yields the ratio
\begin{equation}
\left[ 
{B_{n+1}\over B_n } \right]  e^{-i(\omega_{n+1}-\omega_n)( t-r_\ast-y)} 
\ .
\label{ratio}\end{equation}
Now we know that \cite{nollert,andersson93} 
\begin{equation}
\omega_{n+1}M \approx \omega_n M - i/4 \ , \quad {\rm as} \quad n\to \infty \ .
\end{equation}
Assuming that the term in the square brackets remains of order unity
(say) it follows that the magnitude of the ratio of successive terms in
the mode sum behaves asymptotically as $\exp[-( t-r_\ast-y)/4]$. This
implies that the sum will surely converge for $t-r_\ast-y \gg 0$.

\begin{figure}
\centerline{\epsfxsize=200pt \epsfbox{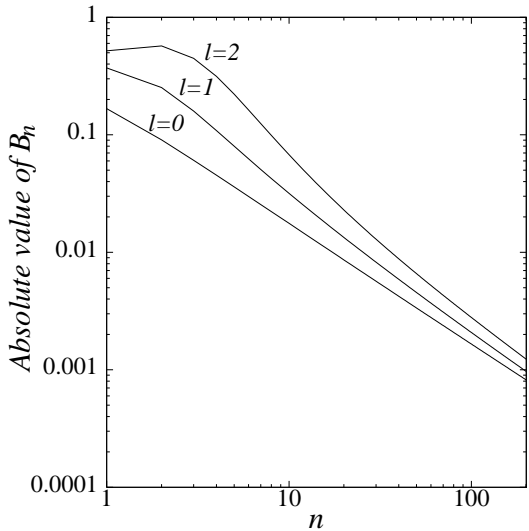}}
\caption{The absolute value of the terms in the mode sum at $t-r_\ast-y=0$ for
$\ell=0,1$ and 2 are shown as a function of the mode index $n$ for the 
first 200 modes. The data is obtained using approximate phase-integral expressions. }
\label{fconv}\end{figure}

But this argument relies on the terms in the square brackets of
(\ref{ratio}) behaving in a certain way. Does it hold in practice?  To
test this we have used the approximate formulae derived in
\cite{andersson95} to obtain $B_n$ for the first 200 modes of a scalar
field (and $\ell=0,1$ and 2). These results are illustrated in
Figure~\ref{fconv}. The data  here is not expected to be very accurate
for the highly damped modes. Nevertheless, the trend is clear. The
magnitude of $B_n$ decreases monotonically for large values of $n$.
Moreover, one finds that successive terms have opposite signs.  This
suggests a stronger convergence than the expected one: The mode sum
will converge also at $t - r_\ast - y = 0$ .

\subsection{Dynamic mode-excitation} 

Previous investigations of this problem \cite{andersson95,sun_price}
were to a certain extent marred by what can be called the ``timing
problem''.  Assume that the initial data consists of a ``mountain''
close to the black hole and a tiny ``pimple'' far away.  It was then
found that the ``pimple'' leads to a much larger mode-excitation than
does the ``mountain''. This is, of course, contrary to our
expectations. Fortunately, it is also wrong.

The ``timing problem'' arises when one tries to associate a given set
of initial data with a constant ``excitation strength'' of each
quasinormal mode. This would be a useful approach for many familiar
oscillating systems, such as a vibrating string. But this approach is
probably only meaningful  when the modes of the system form a complete
set. In the black-hole case it is well known that the quasinormal modes
are not complete (one must also take account of the branch cut
integral).  As we shall see, it makes more sense to consider the
quasinormal-mode excitation as a dynamic process.

For example, one would not expect the quasinormal modes to be
(considerably) excited until after the relevant feature in the inital
data has scattered off the potential barrier that surrounds the hole.
In our previous example this means that the mode-excitation that arises
because of the ``mountain'' in the data will be relevant earlier than
that associated with the ``pimple''. Roughly, the modes should be
excited when the relevant data reaches the peak of the effective
potential ($r_\ast\approx 0$).

It is not very difficult to obtain a dynamic description of the
mode-excitation. In fact, we  need only ensure that the Green's
function respects causality. The basic ``mistake'' of the previous
studies \cite{andersson95,sun_price} was to perform the integral over
the product of the initial data and the Green's function
$\hat{G}(r_\ast,y,\omega)$ before inverting the integral transform. If
one instead uses (\ref{Geqn}) the confusion of the ``timing problem''
can be avoided.

To ensure causality we must ensure that the Green's function vanishes
when $t-r_\ast+y>0$. An observer (at $r_\ast$) should simply not see
anything until after a signal travelling at the speed of light has been
able to reach him/her from the relevant source point ($y$).  In the
evolution equation (\ref{Geqn}) this translates into a {\em lower limit
of integration} $y=r_\ast-t$.
 
For the quasinormal modes, one can 
also argue for the use of an upper limit of integration:
If we use (\ref{greenapp}) it is clear that the quasinormal-mode 
part of the signal arises from 
\begin{equation} 
G(r_\ast,y,t) \sim -{1 \over 4\pi i} \int_C {1 \over \omega}
{ A_{\rm out} \over A_{\rm in} }
e^{-i\omega(t-r_\ast-y)} d\omega \ .
\end{equation}
Intuitively, one would not expect it to be meaningful to close the
integration contour $C$ in the lower half-plane unless $t-r_\ast-y>0$.
At earlier times it seems unlikely that the contribution from the
necessary high-frequency arcs will vanish. We will consider this issue
in more detail in section V.  For now we are content to deduce that
this introduces an {\em upper limit of integration} $y=t-r_\ast$ in the
evolution equation (\ref{Gevol}).

\begin{figure}
\centerline{\epsfxsize=200pt \epsfbox{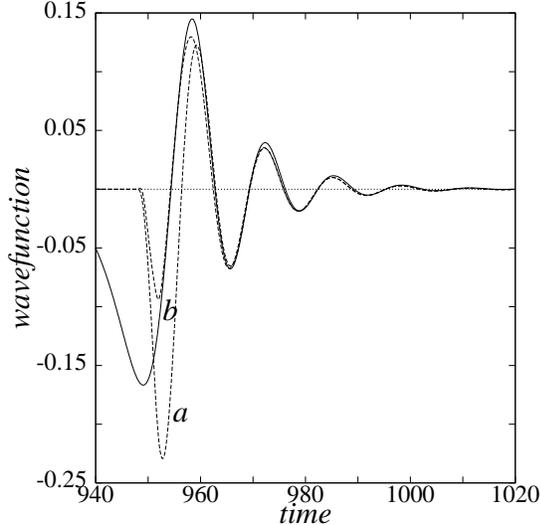}}
\caption{Comparing the field obtained through evolving the
 scalar wave equation to the approximate contribution from the
quasinormal modes. The graph shows data for $\ell=2$. 
The solid line represents the true scalar wave, while
the two dashed lines are for a) the slowest damped quasinormal mode
and b) the sum of the first six modes.}
\label{fring}\end{figure}

As a simple example of the implications of this discussion
we consider the static initial data
\begin{eqnarray}
u(r_\ast,0) &=& \exp[-0.05(r_\ast/M-400)^6] \ , \\
\partial_t u(r_\ast,0) &=& 0 \ .
\label{data}\end{eqnarray}
It is straightforward to multiply this data with $G^Q(r_\ast,y,t)$ from
(\ref{Gmodes}) and then integrate (numerically) from $y=r_\ast-t$ to
$y=t-r_\ast$. The result of this calculation is displayed in
Figure~\ref{fring}. Here we have assumed that the observer sits at
$r_\ast = 500M$. This means that, according to the ``dynamical''
description the modes  should not be present in the  signal before
$t\approx 950M$.

From the data presented in Figure~\ref{fring} we can conclude
that our approximation for the quasinormal mode ringing is quite
accurate. It is conceivable that the remaining discrepancies at later
times would disappear if we used the true mode-functions instead of the
``asymptotic approximation''.  The most interesting part of
Figure~\ref{fring} concerns the early times. It is notable how nicely
the idea of a ``dynamic'' mode excitation works in practice. This
allows us to discuss the relevance of the high-order modes in more
detail than what has been possible before
\cite{leaverprd,andersson95,sun_price}. Since they are rapidly damped,
one would expect the highly damped modes to be relevant only at early
times. As can be seen in Figure~\ref{fring} this is, indeed, the case.
While the slowest damped mode represents the signal well after (say)
$t\approx 965M$, a much better approximation (valid from (say)
$t\approx 953M$) is obtained by using the sum of the first six modes.
Should we include further modes in the sum this trend continues, but it
becomes hard to distinguish any improvement.  This evidence clearly
supports the notion that the mode-sum converges for all times after
$t-r_\ast-y=0$.

\section{The late-time power-law tail}

It is by now well-known that the quasinormal-mode ringing 
is followed by a power-law tail at late times.
This feature was first found in the seminal work of  Price 
\cite{price1,price2}. Physically, the tail arises because
of backscattering off the slightly curved spacetime in the
region far away from the black hole \cite{thorne}. This means that the 
tail will not depend on the exact nature of the central object. 
Thus, a neutron star of a certain mass will give rise to
the same tail as would a black hole of the same mass.

Mathematically, it has been demonstrated that the power-law tail can be
associated with the branch cut in the complex-frequency Green's
function \cite{leaverprd}. Using this fact, the exact form of the
leading order tail has been calculated in different ways.  But some
questions still remain. The most important one concerns the black-hole
response at intermediate times --- after the quasinormal modes have
died away, but before the leading tail term accurately represents the
evolution. Is it possible to extend  existant calculations in such a
way that one can approximate the evolution also at these intermediate
times? Previous calculations are also somewhat involved; it would be
nice to have a simpler and more direct calculation of the tail effect
for black holes.

One can argue that the late-time behaviour should follow from the
low-frequency contribution to the Green's function.  Basically, the
effective black-hole potential in (\ref{rweq}) will be so small for
large values of $r$ that only low-frequency waves will be affected by
it.  Hence, a low-frequency approximation to the black-hole equation
will be useful.  In this section we obtain such an approximation, and
use it to study the detailed behaviour of the power-law tail.

\subsection{A low-frequency approximation}

Let us begin by introducing a new dependent variable 
\begin{equation}
\hat{u}_\ell = \left( 1 - {2M \over r} \right)^{-1/2} \psi \ .
\end{equation}
Then equation (\ref{rweq}) for the scalar field becomes
\begin{equation}
\left[ {d^2 \over dr^2} + \left( 1 - {2M \over r} \right)^{-2} \left\{
\omega^2 - \left(1-{2M\over r} \right){\ell(\ell+1) \over r^2} 
+ {M^2 \over r^4} \right\} \right] \psi = 0 \ .
\label{psieq}\end{equation}
As already mentioned above, it is easy to argue that 
we only need a large $r$ approximation to account
for the low-frequency response. Thus,  we  expand 
(\ref{psieq}) as a power series in $r/M$. This leads to
\begin{equation}
\left[ {d^2 \over dr^2} + \omega^2 + {4M\omega^2 \over r}
- {\ell(\ell+1) \over r^2}  \right] \psi \approx 0 \ .
\label{loweq}\end{equation}
The question is whether this is a useful approximation \cite{footn}.
That is, is the assumption that we only need large $r$ justified?
Fortunately, this is a simple thing to test. From (\ref{loweq}) we find
that the outer turning point of classical motion is located at
\begin{equation}
r_{tp} = 2M \left\{ \sqrt{ {\ell(\ell+1) \over 4M^2\omega^2} - 2} - 1
\right\} \ .
\end{equation}
At $r_{tp}$ one would expect a wave with frequency $\omega$ to be
scattered by the effective potential.  Since $r_{tp}\to +\infty$ as
$\omega M \to 0$ it is clear that the approximate equation
(\ref{loweq}) can be used with confidence for low frequencies.

Let us now introduce \cite{futterman}
\begin{equation}
\psi = \left( {r\over M} \right)^{\ell+1} e^{i\omega r} \phi(z) \ ,
\end{equation}
where $z = -2i\omega r$.
Then, it follows that $\phi$ should be a solution to the confluent 
hypergeometric equation
\begin{equation}
\left[ z {d^2 \over dz^2} + (2\ell+2 -z) {d \over dz} - 
(\ell+1 -2i\omega M) 
\right] \phi = 0 \ .
\label{hyperge2}\end{equation}
It also follows that two basic solutions to the black-hole problem 
can be written (remember that $r>>2M$)
\begin{equation}
\hat{u}_\ell^{\rm in} = A \left({r\over M}\right)^{\ell+1} 
e^{i\omega r} M (\ell+1-2i\omega M, 2\ell+2, -2i\omega r) \ ,
\end{equation}
and
\begin{equation}
\hat{u}_\ell^{\rm up} = B \left({r\over M}\right)^{\ell+1} 
e^{i\omega r} U (\ell+1-2i\omega M, 
2\ell+2, -2i\omega r) \ ,
\end{equation}
where $A$ and $B$ are normalisation constants. The functions $M(a,b,z)$
and $U(a,b,z)$ represent the two standard solutions to the confluent
hypergeometric equation  \cite{abra-stegun}.

One nice feature of this approximation is that it is obvious that there
will be a cut in $\hat{u}_\ell^{\rm up} $.  We simply use the standard
result \cite{abra-stegun} that, for $n$ an integer, 
\begin{eqnarray}
U(a, n+1, z) &=& { (-1)^{n+1} \over n!\Gamma(a-n) } M(a, n+1, z) \ln z
+ \nonumber \\
&+& \mbox{single-valued terms} \ .
\label{loginu}\end{eqnarray}
From this it follows immediately that
\begin{eqnarray}
U(a, n+1, ze^{2\pi i}) &=& U(a, n+1, z)  \nonumber \\
&+& 2\pi i { (-1)^{n+1} \over n!
\Gamma(a-n) } M(a, n+1, z) \ .
\label{cutu}\end{eqnarray}
We will now use this result to evaluate the effect of the branch cut in 
the Green's function.

\subsection{The late-time tail}

We have seen that there will necessarily be a branch cut in the
black-hole Green's function $\hat{G}(r_\ast,y,\omega)$ (or more
specifically, in the solution $\hat{u}_\ell^{\rm up} $).  To arrive at
the detailed power-law tail we need to consider the effect of this cut.
It is easy to see that the contribution from the branch cut follows
from the integral [cf. (\ref{greent})]
\begin{eqnarray}
&& G^C(r_\ast, y, t) = \nonumber \\
&& {1\over 2\pi} \int_0^{-i\infty} 
\hat{u}_\ell^{\rm in}(y,\omega) \left[ 
{ \hat{u}_\ell^{\rm up}(r_\ast,\omega e^{2\pi i}) \over
W(\omega e^{2\pi i}) } \right. \nonumber \\
&& \quad \quad - \left. { \hat{u}_\ell^{\rm up}(r_\ast,\omega) \over
W(\omega) }   \right] e^{-i\omega t} d\omega \ .
\end{eqnarray}
After some straightforward steps this can be rewritten as
\begin{eqnarray}
&& G^C(r_\ast, y, t) =  \nonumber \\
&&{1\over 2\pi} \int_0^{-i\infty} 
\hat{u}_\ell^{\rm in}(y,\omega) \hat{u}_\ell^{\rm in}(r_\ast,\omega) 
 \times \nonumber \\
&& \quad \quad \times { W[\hat{u}_\ell^{\rm up}(r_\ast,\omega), 
\hat{u}_\ell^{\rm up}(r_\ast,\omega e^{2\pi i})] \over  
W(\omega e^{2\pi i}) 
W(\omega) } e^{-i\omega t} d\omega \ .
\end{eqnarray}

Using the low-frequency approximation that was described in 
the previous section we can  show that
\begin{equation}
W(\omega)  =
(-1)^{-\ell-1} iAB 
{ (2\ell+1)! (2\omega)^{-2\ell-1} \over \Gamma(\ell+1-2i\omega M) } \ ,
\end{equation}
and since there is no cut in $\hat{u}_\ell^{\rm in} $ 
it is easy to see from (\ref{cutu}) that 
\begin{equation}
W(\omega e^{2\pi i}) = W(\omega) \ .
\end{equation}
To build the Green's function we also need
\begin{eqnarray}
&&W[\hat{u}_\ell^{\rm up}(r_\ast,\omega), 
\hat{u}_\ell^{\rm up}(r_\ast,\omega e^{2\pi i})] = \nonumber \\ 
&& \quad B^2 { (-1)^{-\ell-1} 2\pi (2\omega)^{-2\ell-1} \over
\Gamma(-\ell-2i\omega M) \Gamma(\ell+1-2i\omega M) } \ .
\end{eqnarray}
If we use these results together with the approximation
\begin{eqnarray}
\hat{u}_\ell^{\rm in} &\approx& A \left({r\over M}\right)^{\ell+1} 
e^{i\omega r} M(\ell+1, 2\ell+2, -2i\omega r) = \nonumber \\
&=& 
A (2\ell+1)!! (\omega M)^{-\ell} \left( { r\over M} \right) 
j_\ell(\omega r) \ ,
\end{eqnarray}
we get
\begin{equation}
G^C(r_\ast, y, t) = 4iMr_\ast y \int_0^{-i\infty} \omega^2 
j_\ell(\omega r_\ast)
j_\ell(\omega y) e^{-i\omega t} d\omega \ .
\label{cutint}
\end{equation}  
To obtain this result we have assumed that the asymptotic approximation
is valid. Specifically, we have replaced $r$ by $r_\ast$. This is not
at all a necessary step, but it simplifies the comparison to our
numerical evolutions of (\ref{waveq}) which were carried out using
$r_\ast$ as independent variable.  Moreover, if $r$ is sufficiently
large we will only introduce a small error.  In the specific case of
(\ref{data}) the error introduced in this way will be smaller than 2.5
\% .

As long as we are only interested in the leading order behaviour at 
very late times,
we can assume that $\omega y < \omega r_\ast <<1$ \cite{leaverprd}. 
Thus, using the standard
power series expansion for $j_\ell(z)$ we arrive at the 
final formula
\begin{equation}
 G^C(r_\ast, y, t) =
 (-1)^{\ell+1} { (2\ell+2)! \over [ (2\ell+1)!!]^2 }  
{ 4M (r_\ast y)^{\ell+1} \over t^{2\ell+3} }  \ .
\label{tail}\end{equation}
This result is identical to that obtained by Leaver \cite{leaverprd}.
This is not surprising since our derivation is just a simplification of
Leaver's approach. Our result also agrees with that of Ching {\em et
al} \cite{ching}, but in that case one must do some additional
calculations to obtain the explicit result for the black-hole case.
 
We have not yet contributed much new information. The greatest merit of
the above derivation is its simplicity.  The origin of the branch cut
is clear and its contribution to the Green's function follows
painlessly.  This is in remarkable contrast to, for example, the work
of Ching {\em et al} \cite{ching}, where the result follows after a
truly involved analysis.  On the other hand, the formulae obtained by
Ching {\em et al.} are valid for a large class of potentials. Hence,
their work shows that the tail-phenomenon is a generic feature of many
problems in wave scattering.

Simplicity is not the sole advantage of the present approach, however.
It turns out that we can easily do the full integral in (\ref{cutint}).
As long as $t-r_\ast-y>0$ we can employ equation (6.626) from
Gradshteyn and Ryzhik \cite{gradshteyn}. This leads to a higher-order
result
\begin{eqnarray}
&& G^C(r_\ast, y, t) =   \nonumber \\
&& M 
\sum_{m=0}^\infty \sum_{n=0}^m { (-1)^{\ell+1} 2^{2-m} (2\ell+2m+2)! \over n! (m-n)!
(2\ell+2n+1)!! (2\ell+2m-2n+1)!!} \times \nonumber \\
&& \quad \quad \times
{ r_\ast^{\ell+2m-2n+1} y^{\ell+1+2n} \over  t^{2\ell+3+2m} } \ ,
\label{hotail}\end{eqnarray}
which is remarkably simple.

\begin{figure}
\centerline{\epsfxsize=200pt \epsfbox{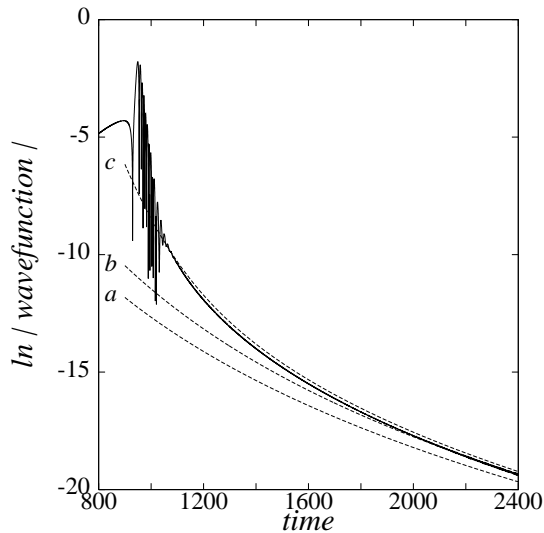}}
\caption{Comparing the field obtained through evolving the scalar wave
equation to the approximate contribution from the power-law tail. The
graph shows data for $\ell=2$ using a logarithmic scale.  The solid
line represents the true scalar wave, while the three dashed lines are
for a) the leading order power-law tail b) the tail approximation
including the first two terms and c) using the first 11 terms in the
tail-sum.}
\label{ftail}\end{figure}

We have thus managed to extend previous work to include higher order
corrections to the power-law tail. The question is to what extent such
a result is useful. Interestingly, it turns out that the higher order
terms play an important role. Essentially they allow us to extend the
validity of the tail-approximation to much earlier times. A typical
result --- obtained for the initial data given in (\ref{data}) --- is
shown in Figure~\ref{ftail}.  Although it is not clear from
Figure~\ref{ftail},  it can be verified that the lowest order tail-term
is a reasonable approximation to the true evolution at very late times.
But the improvement achieved by including the first two terms in
(\ref{hotail}) is impressive. It is also clear from Figure~\ref{ftail}
that if one includes several terms in (\ref{hotail}) one arrives at an
approximation that takes over from the quasinormal-mode ringing in a
natural way.

\begin{figure}
\centerline{\epsfxsize=200pt \epsfbox{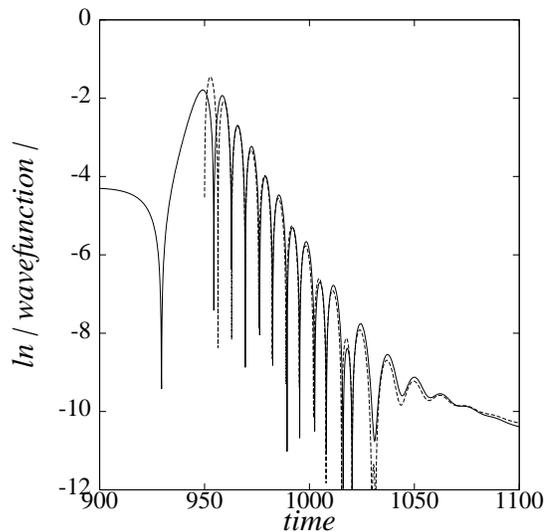}}
\caption{Comparing the field obtained through evolving the scalar wave
equation to the approximate contribution from the slowest damped
quasinormal mode and the first 11 terms in the sum for the power-law
tail.  The graph shows data for $\ell=2$ using a logarithmic scale.
The solid line represents the true scalar wave, while the dashed line
represents the approximation. It is easy to distinguish effects of
interference between the mode and the tail-terms (whenever the ringing
differs from pure expontential damping at a constant oscillation
frequency). } \label{fhotail}\end{figure}

Figure~\ref{ftail} shows that there will be a considerable time window
in which the contributions from the quasinormal-modes and the higher
order tail are of the same order of magnitude.  It seems reasonable to
assume that this may lead to interference effects that can be
distinguished in the evolution of the field.  That such effects are
present is clear from Figure~\ref{fhotail}.  It is also clear that a
combination of the quasinormal modes and the higher order tail-sum
provides a good representation of the evolution throughout the
transition between the typical times when either term dominates the
signal. Thus the approximations discussed so far can be combined to
estimate the signal for all times $t-r_\ast-y>0$.

\section{High frequencies}

In the complex-frequency approach it is typically assumed that the
contribution from the required arcs at $\vert \omega \vert \to \infty$
is irrelevant. That way, the original contour-integral can often be
replaced by a mode-sum such as (\ref{Gmodes}). In the black-hole
problem the situation is, of course, complicated by the  branch cut in
the Green's function.  But as we have seen, the contribution from this
cut can readily be approximated (at least at late times).  To complete
the study of the initial value problem we now focus our attention on
the high-frequency problem.

Although it is easy to give a handwaving argument that suggests that
the large-frequency arcs should not contribute significantly to the
Green's function (when $\vert \omega M\vert$ is very large one would
not expect the details of the effective potential to matter much; a
high-frequency wave will propagate almost as in flat space) the issue
has not been studied in much detail previously. A better understanding
of the high-frequency problem is useful for several reasons: i) We
obviously should confirm that our expectations of a vanishing
contribution to the Green's function holds.  ii) The high frequencies
may hold the key to the response of the black hole at early times
\cite{leaverprd}.  Hence, it is plausible that an understanding of the
behaviour for high frequencies will yield a handle on the inital part
of the signal that reaches an observer.

\subsection{An approximation for high frequencies}

An approximation that is relevant for high frequencies can
be obtained in the following way :
When $|\omega M|$ becomes large the equation that governs the scalar field
(\ref{psieq})  limits to
the confluent hypergeometric equation \cite{liu96}
\begin{equation}
\left[ z {d^2 \over dz^2} + (2\mu -z) {d \over dz} - (\mu -2i\omega M) 
\right] \phi = 0 \ .
\label{hypergeo}\end{equation}
Here we have used
\begin{equation}
\psi = \left({r\over 2M} -1 \right)^\mu \exp\left[ i\omega(r-2M) \right]\phi \ ,
\end{equation}
\begin{equation}
z = -2i\omega(r-2M) \ ,
\end{equation} 
and $\mu$ is given by
\begin{equation}
\mu = {1\over 2}  - 2 i \omega M \ .
\end{equation}

As in the case of low frequencies it is easy to use standard formulae from \cite{abra-stegun} 
and show that 
the two basic solutions that we require to build the
black-hole Green's function can be written
\begin{eqnarray}
&&\hat{u}_\ell^{\rm in} = \left( {r\over 2M} \right)^{1/2} 
\left( {r \over 2M} - 1 \right)^{-2i\omega M} \times \nonumber \\
&& \times M(1/2-4i\omega M , 
1 - 4i \omega M , -2i\omega (r-2M) ) e^{i\omega(r-4M)} \ ,
\label{solu1}\end{eqnarray}
and
\begin{eqnarray}
&&\hat{u}_\ell^{\rm up} = (-4i\omega M)^{1/2-4i\omega M} \left( {r \over 2M}
\right)^{1/2} \left( {r\over 2M} -1 \right)^{-4i\omega M} \times \nonumber
\\
&& \times U(1/2 - 4i \omega M , 1 - 4i \omega M, 
-2i\omega(r-2M) ) e^{i\omega r_\ast} \ .
\label{solu2}\end{eqnarray}

From the asymptotic behaviour of the confluent hypergeometric
functions \cite{abra-stegun} it follows that (in the right 
half of the complex $\omega$-plane)
\begin{equation}
A_{\rm out} = { \Gamma(1-4i\omega M) (4i\omega M)^{-1/2+4i\omega M} 
e^{-4i\omega M} \over \sqrt{\pi} } \ ,
\end{equation}
and
\begin{equation}
A_{\rm in} = { \Gamma(1-4i\omega M) (4i\omega M)^{-1/2} e^{i\pi/2} 
\over \Gamma(1/2-4i\omega M) } \ .
\label{ain}\end{equation}
After using large argument approximations for the $\Gamma$
functions we  get
\begin{equation}
A_{\rm out} \approx i \sqrt{2} e^{-4\pi \omega M} \ ,
\end{equation}
and
\begin{equation}
A_{\rm in} \approx 1 \ .
\end{equation}
Hence, a high-frequency approximation to the 
``reflection coefficient'' of the black hole is
\begin{equation}
{\cal R} = \vert {A_{\rm out} \over A_{\rm in} } \vert^2 \approx e^{-8\pi 
\omega M} \ .
\end{equation}
This result agrees with our expectations (from for example the
WKB method): For very large frequencies the reflection caused by
the black-hole potential barrier will be exponentially small.

We can also use this approximation to approximate the very high
overtones of the black hole. Recall that the quasinormal modes follow
from $A_{\rm in} = 0$. Then, it is trivial to use (\ref{ain}) and show
that  modes should be located at \cite{liu96}
\begin{equation}
\omega_n M = - {i\over4} \left( n + {1\over 2} \right) \ . 
\end{equation}
This approximation yields the correct damping rate for the high modes,
but it fails to reproduce the small constant real part that each mode
should have \cite{nollert,andersson93}.

\subsection{High-frequency Green's function}

We now want to use these results to discuss the possible contribution
to the black hole Green's function from the required arcs at $\vert
\omega M \vert \to \infty$. To do this we assume that the asymptotic
approximation from section IID is appropriate (we will discuss
alternatives to this later). Then we clearly get (in the right
half-plane)
\begin{equation}
\hat{G}^\infty(r_\ast,y,\omega) \approx -{ 1 \over 2i\omega} 
\left[ e^{i\omega(r_\ast-y)} + i \sqrt{2} e^{-4\pi \omega M} 
e^{i\omega(r_\ast+y)} \right]  \ . 
\label{highg}\end{equation}
For obvious reasons it makes sense to study the two terms in the
square brackets separately. 

For each term we require an integral 
of form [ cf. (\ref{greent})]
\begin{equation}
I = \int_C {f(\omega) \over \omega}  e^{-i\omega \tau}d\omega
\ ,
\end{equation}
where $C$ is a large-frequency quarter circle in either the upper or the
lower right half of the $\omega$-plane  (the calculation here must 
be complemented by similar formulae for the left half-plane). Now, as long as
\begin{equation}
{f(\omega) \over \omega} \to 0 \ , \quad {\rm as} \quad \vert \omega \vert \to \infty \,
\end{equation}
(which is certainly true here) the integral $I$ vanishes i) in the upper half-plane
for $\tau<0$, and ii) in the lower half-plane for $\tau>0$ (Jordan's lemma).

For the specific case of the black-hole Green's function this means that
\begin{equation}
G(r_\ast,y,t) = 0 \ , \quad {\rm for} \quad \left\{ \begin{array}{ll} 
& t-r_\ast+y < 0 \ , \\
& t-r_\ast-y > 0 \ . \end{array} \right. 
\end{equation}
This is a nice result because it shows that a combination of the
quasinormal modes and the contribution from the branch cut in
$\hat{G}(r_\ast,y,\omega)$ should form a complete description of the
wave evolution after $t-r_\ast-y=0$. This confirms the result of the
previous section. It also shows that the evolution should be causal: No
signal will reach the observer before $t-r_\ast+y=0$. But the situation
for the intermediate times is not clear.

\subsection{Approximating the signal at earlier times}

We have seen (cf. Figure~\ref{fhotail}) that the combination of the
quasinormal modes and the higher order tail sum provides a complete
description of the evolution of an initial wave field after a certain
time. The only remaining question is whether we can approximate the
evolution adequately also at earlier times. Unfortunately, this turns
out to be  harder than what one might expect.

To discuss this issue we consider the high-frequency Green's function
(\ref{highg}). For intermediate times $r_\ast-y\le t \le r_\ast+y$ one
would naively want to use a contour in the lower half-plane for the
first term in (\ref{highg}) while at the same time using a contour in
the upper half-plane for the second term. But although consistent with
our discussion of quasinormal modes in section III this is not a useful
approach. The main reason is that we should not treat the two terms in
(\ref{highg})  independently. This does not mean that our previous
discussion is flawed: Our results still hold because after
$t-r_\ast-y>0$ one would certainly want to close both integration
contours in the lower half-plane. But problems arise when we want to
consider earlier times.

\begin{figure}
\centerline{\epsfxsize=200pt \epsfbox{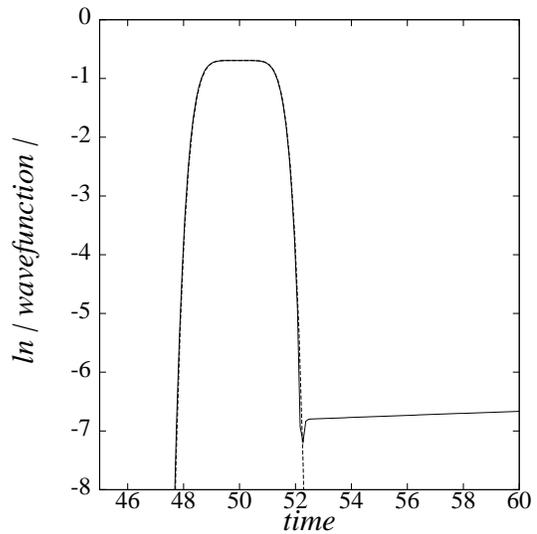}}
\caption{Comparing the field (for $\ell=2$) obtained through evolving
the scalar wave equation to an approximation that corresponds to
propagation in flat space. The initial signal that reaches the observer
is shown. After the, essentially unchanged, initial shape follows a
tiny wake. This wake will be more pronounced for data that has support
closer to the black hole.  The solid line represents the true scalar
wave, while the dashed line represents the approximation. }
\label{fearly}\end{figure}

Before discussing this problem in more detail it may be helpful to
illustrate the features in the evolution that we want to describe. In
Figure~\ref{fearly} we show the initial signal that reaches the
observer  (at $t-r_\ast+y=0$) in the case of (\ref{data}).  It is clear
that this signal is more or less the direct transmission that one would
expect in flat space. But there is one important difference that is
hard to distinguish in Figure~\ref{fearly}:  After the initial pulse
follows a tiny wake. A description of the early time behaviour must
yield this effect. Another part of the signal that one would like to
describe is the ``reflected'' pulse that reaches the observer slightly
before $t-r_\ast-y=0$, {\em i.e.,} before the onset of quasinormal
ringing in Figure~\ref{fring}.  

So why is this problem difficult? Basically, there are two possible
routes and both  forces us to deal with formally singular terms. The
first possibility is to close the integration contour in the upper
half-plane at all times before $t-r_\ast-y=0$. If we do so it is clear
that the first term in (\ref{highg}) will be singular, but all other
contributions to the Green's function will vanish. The second option is
to close the contour in the lower half-plane as early as
$t-r_\ast+y=0$. Then we find that there will be three divergent terms
(that must balance each other in some magic way): We know that the
quasinormal-mode sum will diverge, and the same is true also for the
integral along the branch cut and the high-frequency arc [the second
term in (\ref{highg})]. Clearly, this second alternative is the less
attractive one and we will take a few steps down the first route here.

Let us  consider only the leading order. As already pointed out, the
only contribution to the Green's function that does not vanish for
$r_\ast-y\le t \le r_\ast+y$ comes from the first term in
(\ref{highg}). Thus, we need to evaluate
\begin{equation}
 G(r_\ast,y,t) = - {1\over 4\pi i} \int_C {e^{-i\omega(t-r_\ast+y)}
 \over \omega} d\omega \ ,
\end{equation}
where the integration contour $C$ is a semi-circle in the upper half-plane.
Now the integrand can be identified as the Laplace-transform of the step-function.
Specifically, we get
\begin{equation}
G(r_\ast,y,t) = {1\over 2} H(t-r_\ast-y) \ .
\label{gflat}\end{equation}
This is, of course, exactly what we would get in flat space. To 
understand the added subtleties of the black-hole
case we must pursue the calculation to higher orders. 

In principle, such a higher order calculation seems possible. One
could, for example, try to express the high-frequency approximations of
the solutions $\hat{u}_\ell^{\rm up}$ and $\hat{u}_\ell^{\rm in}$ from
section VA as power series in  $\omega^{-n}$. That should yield higher
order corrections to (\ref{gflat}). Unfortunately, it seems as if one
would have to keep a large number  of  terms (all?) in the resultant
expression to get a useful answer. There may also be less obvious
complications.  Hence, we will have to return to this problem in the
future.
 
\section{Extending the present work}

To conclude this paper it is meaningful to discuss how the present
work  --- for a massless scalar field in the Schwarzschild geometry ---
can be adapted to other, physically more interesting, cases.

\subsection{Other perturbing fields}

The simplest extension of the present work regards other perturbing
fields.  It is, in fact, trivial to show that all our results carry
over also to electromagnetic and gravitational waves in the
Schwarzschild background. Both the discussion of the tail effect in
section IV and the high-frequency discussion in section V are valid
also for these other test-fields. As regards the quasinormal modes, the
characteristic frequencies (and the coefficients $B_n$ in
(\ref{Gmodes})) will be different for other fields, but the approximate
phase-integral expressions \cite{andersson95} that we used to evaluate
(\ref{Gmodes}) can be used also for electromagnetic and gravitational
perturbations.

\subsection{Without the asymptotic approximation}

Throughout this paper we used the asymptotic approximation from section
IID to simplify the calculations. This clearly restricts the initial
data in an unneccessary way. Fortunately, it is not (formally)
difficult to generalize our results in such a way that they hold also
for more general data.  First we note that the asymptotic approximation
was never used (apart from in the replacement of $r$ by $r_\ast$ in
(\ref{cutint})) in the derivation of the tail-expressions.
Consequently, these results remain unchanged also for general initial
data. For the quasinormal modes, we must use approximations of the
corresponding eigenfunctions that remain valid for all $r$. That this
can be done has already been demonstrated by Leaver \cite{leaverprd}.
The dynamic mode-excitation from  section IIIC becomes more difficult
to introduce in the general case. With the asymptotic approximation it
was easy to find a time before which it would be meaningless to use
integration contours in the lower half of the $\omega$-plane. In the
general case, this specific time might not be so easy to define. This
issue is intimately related to the high-frequency problem. It is clear
that the asymptotic approximation is crucial for the discussion in
section VB. In the general case one should use large-$\omega$
asymptotics for (\ref{solu1}) and (\ref{solu2}) that do not at the same
time assume a large value of $r$. Such asymptotic expressions are not
standard but they should be possible to derive \cite{rachel}. These new
approximations should indicate at what time it is sensible to use
integration contours in the lower half of the $\omega$-plane, and thus
imply an initial time for the excitation of quasinormal modes.
Consequantly, it seems likely that all the ideas presented in this
paper can be useful also in the general case.

\subsection{Rotating black holes}

We finally turn to the problem for rotating black holes. When the black
hole is rotating both the angular  and the radial functions that are
required to describe a perturbation are frequency-dependent. In a
numerical evolution one would therefore not decouple the corresponding
equations, and thus have to deal with a two-dimensional problem. The
analysis of the Kerr problem is, in general, far more complicated than
the present one. But in principle one would expect all the ideas
discussed in this paper to be useful also for rotating holes.

For example, the construction of the black hole Green's function should
be analogous to that in section II. Of course, the angular dependence
would have to be included in equations like (\ref{Gevol}). Then the
quasinormal modes are defined exactly as in the Schwarzschild case. The
main difference is that each Schwarzschild mode splits into $2\ell+1$
distinct ones (for different values of $m$) because of the rotation of
the black hole \cite{leaverprs}.

As regards the late-time power-law tail, the generalization to Kerr
also seems straightforward. Since the issue of tails in the geometry of
a rotating black hole has recently been analyzed through numerical
evolutions by Krivan, Laguna and Papadopoulos \cite{laguna} it may be
worthwhile to discuss this in somewhat more detail here.

As was first shown by Teukolsky, the equations that govern a small
perturbation of a rotating black hole can be written \cite{teukolsky}
\begin{eqnarray}
&& \Delta^{-s} {d \over dr} \left( \Delta^{s+1} {dR \over dr} \right) \nonumber \\
&& \quad + 
\left[ { K^2 - 2is(r-M)K \over \Delta} + 4is\omega r - \lambda \right] R = 0
\ , 
\label{teukrad}\end{eqnarray}
and 
\begin{eqnarray}
&&{1 \over \sin \theta} {d \over d\theta} \left( \sin\theta {dS \over d\theta}
\right) + \left[ a^2\omega^2\cos^2 \theta - {m^2 \over \sin^2\theta} 
-2 a \omega s \cos \theta \right. \nonumber \\
&& \left. - {2ms\cos \theta \over \sin^2 \theta} - s^2\cot^2\theta + 
E - s^2 \right] S = 0 \ .
\label{teukang}\end{eqnarray}
Here we have introduced
\begin{equation}
K \equiv (r^2 + a^2)\omega - am \ ,
\end{equation}
and
\begin{equation}
\lambda \equiv E - s(s+1) + a^2\omega^2 - 2am\omega \ .
\end{equation}
Here $a\le M$ is the rotation parameter of the black hole, and $s$ is
the spin-weight of the perturbing field. The solutions to the angular
equation (\ref{teukang}) are generally referred to as ``spin-weighted
speroidal harmonics''.

Let us now adopt the approach of section IV and approach the Kerr
problem for low frequencies. It is sufficient to consider $s\ge 0$ (the
results for $s=-1,-2$ can be deduced via the Teukolsky-Starobinsky
identities \cite{cbook}), and we can also use the fact that
\cite{seidel}
\begin{equation}
E = \ell(\ell+1) + s^2 + \sum_{n=1}^\infty f_n (a\omega)^n \ .
\end{equation}
Then we expand (\ref{teukrad}) for large $r$ and find
\begin{eqnarray}
&& {d^2 \psi \over dr^2} + \Bigl[ \omega^2 + {4M\omega^2 + 2is\omega \over r} 
\nonumber \\
&& - { \ell(\ell+1) + s^2 - 12M^2\omega^2 + 2am\omega - 2isM\omega \over r^2}\Bigr] \psi = 0 \ ,
\label{kerrlow}\end{eqnarray}
where the new dependent variable $\psi$ is defined by $R=\Delta^{-(s+1)/2} \psi$.
From this equation two things follow immediately: i) For low
frequencies we
can neglect all frequency dependent terms in the factor of $1/r^2$. ii)
For scalar waves ($s=0$) we reproduce the Schwarzschild equation
(\ref{loweq}).  Hence, the leading-order tail calculation in section IV
must remain valid also for scalar waves in the Kerr background. This
confirms the main result of Krivan, Laguna and Papadopoulos
\cite{laguna}. But what about fields with non-zero $s$? Well, from
(\ref{teukang}) and (\ref{kerrlow}) follows that we can neglect all
terms that depend on the rotation of the black hole. This shows that it
is sufficient to analyze the tail effect in the Schwarzschild limit.
The fact that the radial equation (\ref{kerrlow}) does not immediately
agree with (\ref{loweq}) need not worry us too much. As Chandrasekhar
has shown \cite{chandrarel}, the $a\to 0$ limit of (\ref{teukrad}) ---
the Bardeen-Press equation \cite{bardeen} --- can easily be transformed
into the Regge-Wheeler equation (\ref{rweq}).  We thus have clear
evidence that the leading order tail result from section IV will hold
also for Kerr black holes.

To derive higher order tail-corrections for Kerr demands a more
involved analysis. The issue is  complicated by the frequency-dependent
coupling between (\ref{teukrad}) and (\ref{teukang}). Similar
difficulties make the high-frequency problem less transparent. Hence,
we will not discuss these problems further here.  What is clear is that
the Kerr problem poses an interesting challenge, and we hope to be able
to discuss it further in the near future.

\section*{Acknowledgements}

This work was supported by NSF (grant no PHY 92-2290) and NASA (grant no NAGW 3874).


\begin{references}
\bibitem{footn} It should be clear that this low-frequency approximation can be used only for $\ell\ne0$. The special case of $\ell=0$ must be considered separately. This has been done by Ching {\em et al.} \cite{ching}.
\end{references}
\end{document}